%% file: kagome_hub6.tex
\documentclass[epj]{svjour}
\usepackage{bm}
\usepackage{braket}
\usepackage{amsmath}
\usepackage{amssymb}
\usepackage[dvips]{color}
\usepackage[dvipsnames]{xcolor}
\usepackage[dvips]{graphics}
\usepackage[dvips]{graphicx}
\usepackage{eepic}
\usepackage{pstcol}
\usepackage{pstricks}%
\usepackage{pst-eps}%
\usepackage{pst-plot}%
\usepackage{pst-node}%

\renewcommand{\i}{{\rm i}}
\newcommand{\e}{{\rm e}}

\begin{document}
\title{Exact Ground States of the Extended Hubbard Model on the Kagom\'e
lattice}
\author{
Masaaki Nakamura\inst{1}
\and 
Satoshi Nishimoto\inst{2,3}
}                     
%
%
\institute{
Department of Physics, Ehime University
Bunkyo-cho 2-5, Matsuyama, Ehime 790-8577, Japan,\and
Department of Physics, Technical University Dresden, 01069 Dresden,
Germany,\and
Institute for Theoretical Solid State Physics, IFW Dresden, 01171
Dresden, Germany
}
\date{Received: date / Revised version: date}
%
\abstract{
 We discuss the exact plaquette-ordered ground states of the generalized
 Hubbard model on the Kagom\'e lattice for several fillings, by
 constructing the Hamiltonian as a sum of products of projection
 operators for up and down spin sectors.  The obtained exact ground
 states are interpreted as N\'eel ordered states on the bond-located
 electrons. We determine several parameter regions of the exact ground
 states, and calculate the entanglement entropy. We examine the above
 results by numerical calculations based on exact diagonalization and
 density-matrix renormalization group methods.
\PACS{
      {71.10.Fd}{Lattice fermion models (Hubbard model, etc.)}   \and
      {71.10.Hf}{Non-Fermi-liquid ground states, electron phase diagrams and phase transitions in model systems}   \and
      {75.10.-b}{General theory and models of magnetic ordering}   \and
      {71.23.An}{Theories and models; localized states}
     } 
} 
\maketitle
\section{Introduction}

The Hubbard model is one of the generic models to describe strongly
correlated electron systems~\cite{Hubbard}.  This model has played
important roles to study magnetism and superconductivity.  However, in
spite of its simplicity, it is difficult to solve this model exactly
except for one dimension or some special cases.  On the other hand,
extended versions of the Hubbard model have also been studied.  The
on-site repulsion of the Hubbard model is due to the matrix elements of
the Coulomb interaction corresponding to the on-site Wannier states, and
other matrix elements are neglected. Therefore, it is worth considering
the effects of these neglected terms as site-off-diagonal
interactions~\cite{Campbell-G-L}.  For these generalized models, exact
results for ferromagnetism and superconducting states have been
discussed~\cite{Simon-A,Strack-V1993,Strack-V1994,Arrachea-A,%
Boer-K-S,Boer-S,Montorsi-C,Kollar-S-V}.

In addition to those, a different type of exact ground state has been
discussed for a one-dimensional system, which is called ``bond N\'eel''
(BN) state~\cite{Itoh-N-M,Nakamura-I_2001,Nakamura-O-I}, by the
projection operator method~\cite{Majumder-G,Affleck-K-L-T} for
multicomponent systems~\cite{Itoh}. The BN state is regarded as a N\'eel
ordered state of bond-located spins.  Furthermore, the concept of the BN
state in one dimension was extended to higher dimensional systems
introducing plaquette states in corner sharing lattices such as the
Kagom\'e lattice~\cite{Nakamura-I_2004}. In this paper, we extend this
argument for the Kagom\'e Hubbard model at 1/3 filling to several
fillings and give numerical verification based on exact diagonalization
and density-matrix renormalization group (DMRG)~\cite{white92}
techniques. We also calculate the entanglement entropy (EE) exactly.

This paper is organized as follows: In Sec.~\ref{sec:method}, we review
the method to construct Hamiltonians with exact ground states in
multicomponent systems.  In Sec.~\ref{sec:1D}, we review the application
of this method to the one-dimensional model discussed in
Ref.~\ref{Itoh-N-M}. In Sec.~\ref{sec:2D}, we apply the analysis to the
Kagom\'e lattice. In addition to the exact result at 1/3-filling
obtained in Ref.~\ref{Nakamura-I_2004}, we also discuss the results at
2/3-filling and at half-filling.  The exact ground states are
numerically confirmed using the exact diagonalization and DMRG
methods. In Sec.~\ref{sec:EE}, we calculate the entanglement entropy.
Finally, we give summary and discussion of the results.

\begin{figure}[h]
 \input{fig1.tex}
 \caption{Examples of lattice structures where generalized Hubbard
 models with exact plaquette-ordered ground states can be constructed:
 (a) the one-dimensional chain and (b) the Kagom\'e lattice. The blue
 and the red plaquettes denote those belong to the groups $\mathcal{A}$
 and $\mathcal{B}$, respectively.}\label{fig:lattices}
\end{figure}
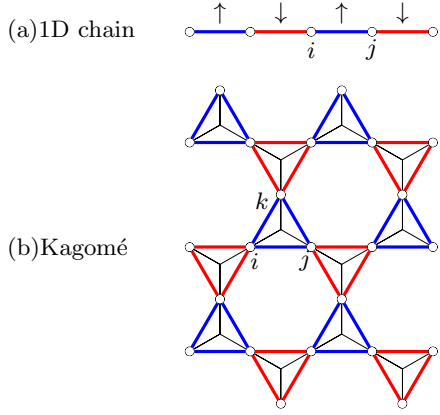

\section{Construction of the Hamiltonian}\label{sec:method}

The method to construct a Hamiltonian with an exact ground state is the
following way~\cite{Itoh}. First, we consider a Hamiltonian given by a
sum of products of projection operators
\begin{equation}
 {\cal H}=\sum_{\alpha} h_{\alpha},\quad
  h_{\alpha}=\sum_{\mu,\nu}\lambda_{\mu\nu}
  R^{(\mu)}_{\alpha\uparrow}R^{(\nu)}_{\alpha\downarrow},
  \quad
  \lambda_{\mu\nu}\geq 0,
  \label{Ham}
\end{equation}
where $\alpha$ denotes the position of one of the unit plaquettes that
cover the lattice.  $R^{(\mu)}_{\alpha\sigma}$ is an operator whose
expectation value is positive semidefinite
$\braket{R^{(\mu)}_{\alpha\sigma}}\geq 0$. This condition is realized,
if $R^{(\mu)}_{\alpha\sigma}$ is given by a product of an operator and
its Hermitian conjugate. Then the expectation value of the Hamiltonian
is also positive semidefinite $\langle {\cal H}\rangle\geq 0$.

Next, we introduce a trial wave function given by a direct product of
up and down spin sectors,
\begin{equation}
  |\Psi(\mathcal{A},\mathcal{B})\rangle=|\Phi_{\uparrow}(\mathcal{A})\rangle
   \otimes|\Phi_{\downarrow}(\mathcal{B})\rangle,\label{state}
\end{equation}
where $\mathcal{A}$ and $\mathcal{B}$ denote two groups of plaquettes
that cover the lattice satisfying
$\mathcal{A}\cup\mathcal{B}=\{\mbox{all lattice sites}\}$.  We require
that the projection operators have the following conditions,
\begin{equation}
 R^{(\mu)}_{\alpha\uparrow}|\Phi_{\uparrow}(\mathcal{A})\rangle
  =R^{(\mu)}_{\beta\downarrow}|\Phi_{\downarrow}(\mathcal{B})\rangle=0,
  \label{method.5}
\end{equation}
where $\alpha\in\mathcal{A}$ and $\beta\in\mathcal{B}$. Therefore, even
if we have
\begin{equation}
 R^{(\mu)}_{\beta\uparrow}
  |\Phi_{\uparrow}(\mathcal{A})\rangle
  \neq 0,\quad
  R^{(\mu)}_{\alpha\downarrow}
  |\Phi_{\downarrow}(\mathcal{B})\rangle\neq 0,
\end{equation}
the eigenvalue of the Hamiltonian for
$|\Psi(\mathcal{A},\mathcal{B})\rangle$ is always zero.  Then, the lower
bound and the upper bound of the energy are coincide, so that
$|\Psi(\mathcal{A},\mathcal{B})\rangle$ turns out to be one of the exact
ground state of this system.

The above argument can be satisfied in corner sharing lattices with the
bipartite structure. The simplest examples is the one-dimensional (1D)
lattice, where the unit plaquette is one bond.  In two dimension (2D),
the Kagom\'e lattice can be covered by two colored triangles
alternatively, as illustrated in Fig.~\ref{fig:lattices}.  These states
can be regarded as the N\'eel ordering on the dual lattice (i.e. the
honeycomb lattice for the Kagom\'e lattice). In three dimension, the
Pyrochlore lattice satisfies these conditions. If the system has a
time-reversal symmetry, its ground state has two-fold degeneracy.

\section{1D model}\label{sec:1D}

We consider the 1D generalized Hubbard model at half-filling and
zero-magnetic field, given by ${\cal H}=\sum_{i\sigma}h_{i,i+1,\sigma}$
with the local bond Hamiltonian,
\begin{align}
 \lefteqn{h_{ij\sigma}=-t\,T_{ij\sigma}
  +\frac{U}{2z}
  (n_{i\sigma}n_{i\bar{\sigma}}+n_{j\sigma}n_{j\bar{\sigma}})
  }\nonumber\\
 &
  +V_{\parallel}n_{i\sigma}n_{j\sigma}+V_{\perp}n_{i\sigma}n_{j\bar{\sigma}}
  \nonumber\\
 &
  +XT_{ij\sigma}(n_{i\bar{\sigma}}+n_{j\bar{\sigma}})
  +\frac{W}{2}\sum_{\sigma'}T_{ij\sigma}T_{ij{\sigma}'},
  \label{local_bond_Ham}
\end{align}
where $\bar{\sigma}$ is the opposite spin of $\sigma$, $z=1$ for the
present 1D case, and periodic boundary conditions are assumed.  We have
defined the hopping and the density operators as
$T_{ij\sigma}\equiv c_{i\sigma}^{\dag}c_{j\sigma}^{}+\mbox{H.c.}$,
$n_{i\sigma}\equiv c_{i\sigma}^{\dag}c_{i\sigma}^{}$.
Note that the bond-bond interaction ($W$) term can be rewritten as
\begin{equation}
 -2W(\bm{S}_i\cdot\bm{S}_{j}+\bm{\eta}_i\cdot\bm{\eta}_{j}
  -{\textstyle\frac{1}{4}}),\label{eqn:W-term}
\end{equation}
where $\bm{S}_i$ and $\bm{\eta}_i$ are the spin and the pseudo spin
operators, respectively.  The components of the pseudo spin operator are
defined by
\begin{equation}
  \eta_i^{+}\equiv(-1)^i c_{i\uparrow}^{\dag}c_{i\downarrow}^{\dag},\ \
  \eta_i^{-}\equiv(-1)^i c_{i\downarrow}c_{i\uparrow},\ \
  \eta_i^{z}\equiv\frac{1}{2}(n_{i\uparrow}+n_{i\downarrow}-1).
  \label{eqn:eta-pairing}
\end{equation}

Now, we introduce the bonding and the anti-bonding operators,
\begin{equation}
 A_{ij\sigma}^{\dag}
  ={\textstyle\frac{1}{\sqrt{2}}}
  (c_{i\sigma}^{\dag}+c_{j\sigma}^{\dag}),\quad
 B_{ij\sigma}^{\dag}
  ={\textstyle\frac{1}{\sqrt{2}}}
  (c_{i\sigma}^{\dag}-c_{j\sigma}^{\dag}).
\end{equation}
The two electron states are given by
$B_{ij\sigma}^{\dag}A_{ij\sigma}^{\dag}
 =c_{i\sigma}^{\dag}c_{j\sigma}^{\dag}$.
These operators on the same bond satisfy the anticommutation relations:
\begin{displaymath}
 \{A_{ij\sigma},A_{ij\sigma'}^{\dag}\}
 =\{B_{ij\sigma},B_{ij\sigma'}^{\dag}\}=\delta_{\sigma\sigma'},\quad
 \mbox{otherwise}=0.
\end{displaymath}
The density operators for the bond operators are given as
\begin{align}
 n_{A\sigma}\equiv&A_{ij\sigma}^{\dag}A_{ij\sigma}
  =\frac{1}{2}(n_{i\sigma}+n_{j\sigma}+T_{ij\sigma}),\\
 n_{B\sigma}\equiv&B_{ij\sigma}^{\dag}B_{ij\sigma}
  =\frac{1}{2}(n_{i\sigma}+n_{j\sigma}-T_{ij\sigma}).
\end{align}
Since we restrict our attention only on the neighboring two sites $i,j$,
we drop these indices from the operators defined above.

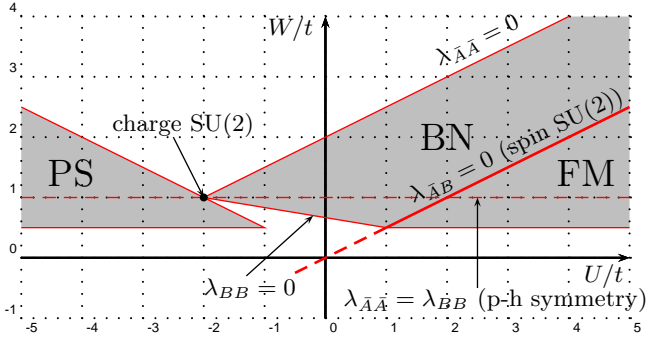
\begin{figure}[t]
 \input{fig2.tex}
 \caption{Phase diagram of the generalized Hubbard chain
 (\ref{local_bond_Ham}) in the $U/2t$-$W/t$ parameter space with
 $t>0$~\cite{Itoh-N-M,Nakamura-O-I}. The parameters are set as $X=t-W$,
 $V_{\parallel}=W$ and $V_{\perp}=U/2$. The shaded regions labeled by
 BN, FM and PS denote bond-N\'eel, ferromagnetic and phase-separated
 states, respectively.}  \label{phase_diagrams_1D}
\end{center}
\end{figure}

As a trial state, we consider the following wave function,
\begin{equation}
 \ket{\Psi_{\sigma}}
  \equiv
  A_{12\sigma}^{\dag}
  A_{23\bar{\sigma}}^{\dag}\cdots
  A_{L-1,L\sigma}^{\dag}
  A_{L,1\bar{\sigma}}^{\dag}
  |0\rangle,
\end{equation}
where $|0\rangle$ denotes a vacuum and $L$ is the number of sites.  This
state is regarded as a N\'eel ordering of the bond-located spins, so
that we call this bond N\'eel (BN) state. There is two-fold degeneracy
given by $|\Psi_{\uparrow}\rangle$ and $|\Psi_{\downarrow}\rangle$. In
order to construct a model with the exact ground state, the local
Hamiltonian $h_{ij}=\sum_{\sigma}h_{ij\sigma}$ should be decomposed by
the projection operators $1-n_{A\sigma}$ and $n_{B\sigma}$ in the
following form,
\begin{align}
h_{ij}&-\varepsilon_0
   =\lambda_{\bar{A}\bar{A}}(1-n_{A\uparrow})(1-n_{A\downarrow})
   +\lambda_{BB}n_{B\uparrow}n_{B\downarrow}\nonumber\\
 &+\lambda_{\bar{A}B}\{(1-n_{A\uparrow})n_{B\downarrow}
   +n_{B\uparrow}(1-n_{A\downarrow})\},
   \label{decomposed_Hamiltonian_1D}
\end{align}
where $\varepsilon_0$ is the ground-state energy per bond. According to
the argument given in Sec.~\ref{sec:method}, for the BN ground state,
the parameters should be chosen as
\begin{equation}
 \lambda_{\bar{A}\bar{A}},\quad \lambda_{\bar{A}B},\quad\lambda_{BB}
  \geq 0.
\label{condition_of_coefficients}
\end{equation}
Comparing Eqs.~(\ref{local_bond_Ham}) and
(\ref{decomposed_Hamiltonian_1D}) (see Appendix~\ref{details}), the
relations among the parameters are obtained as
\begin{equation}
 V_{\perp}=\frac{U}{2},\quad V_{\parallel}=W,\quad X=t-W.
\end{equation}
The coefficients in Eq.~(\ref{decomposed_Hamiltonian_1D}) are identified
as follows,
\begin{align}
 \lambda_{\bar{A}\bar{A}}
  =&\frac{U}{2}-W+2t,\label{l_AA1}\\
 \lambda_{\bar{A}B}
  =&-\frac{U}{2}+W,\label{l_AB1}\\
 \lambda_{BB}
  =&\frac{U}{2}+3W-2t,\label{l_BB1}\\
 \varepsilon_0
  =&\frac{U}{2}.
\end{align}
From Eqs.~(\ref{condition_of_coefficients}), (\ref{l_AA1}),
(\ref{l_AB1}) and (\ref{l_BB1}), we obtain the parameter space of the
exact BN ground state as shown in Fig.~\ref{phase_diagrams_1D}.  Note
that the BN state appears only for $t>0$ region.

The property of the BN state can be investigated based on the
matrix-product method. According to Ref.~\ref{Itoh-N-M}, both
charge-charge and spin-spin correlation functions vanish except for
those of the nearest sites which indicates the existence of the charge
and the spin gaps. On the other hand, the bond-located spin correlation
exhibits a long range order. We can also calculate elementally
excitation spectrum using the matrix-product method as a variational
approach~\cite{Nakamura-O-I}.

In the present one-dimensional model at half-filling, we can discuss not
only the BN state but also the ferromagnetic (FM) and the
phase-separated (PS) states. The last term of
Eq.~(\ref{decomposed_Hamiltonian_1D}) stabilizes the fully polarized FM
state for $\lambda_{\bar{A}B}<0$.  Similarly, the PS state where the
system is separated into a domain of doubly occupied sites and a vacuum,
is stabilized when $\lambda_{\bar{A}\bar{A}}+\lambda_{BB}<0$, neglecting
the surface energy.  As shown in Fig.~\ref{phase_diagrams_1D}, the FM
and the PS states appear in the $U/2t$-$W/t$ parameter space
symmetrically in the positive- and in the negative-$U$ regions,
respectively. This is consistent with the fact that the $W$ term is the
ferromagnetic exchange interactions of the spins and the pseudo spins
(\ref{eqn:W-term}), and the PS state is regarded as the FM state of the
pseudo-spin space.  The condition $W/t\geq 1/2$ for the FM and the PS
phases is not clearly obtained in the present argument.  To obtain this
condition, we need to introduce three-types of $R$
operators~\cite{Nakamura-O-I}.

The phase boundary of the BN and the FM states $\lambda_{\bar{A}B}=0$
corresponds to the SU(2) symmetry in the spins $V_{\parallel}=V_{\bot}$,
so that the ground state is highly degenerate.  The system undergoes a
first-order phase transition at this level-crossing point. When $W/t=1$
($X=0$), the system has the particle-hole symmetry. At
$(U/2t,W/t)=(-1,1)$, the system has the SU(2) symmetry in the
pseudo-spin space, so that the BN, the PS and the $\eta$-paring states
are degenerate.  The other lines which separate shaded and non-shaded
regions in Fig.~\ref{phase_diagrams_1D} do not necessarily mean phase
boundaries.


\begin{figure}
\begin{center}
\includegraphics[width=7cm]{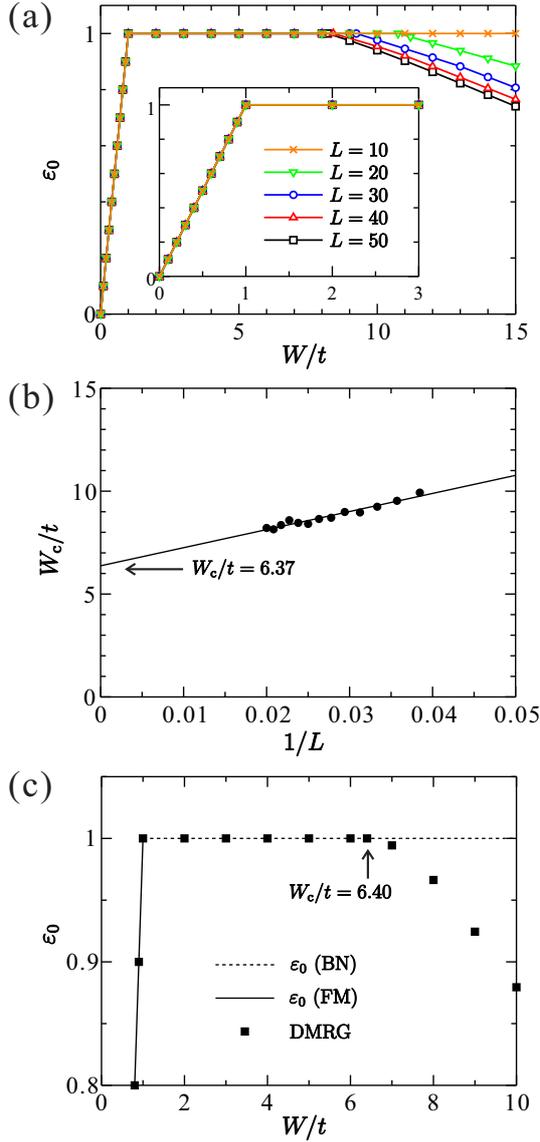}
\caption{(a) Ground-state energy per site as a function of $W/t$ for the
1D model at $U/t=2$, obtained by the exact diagonalization and DMRG with
finite-$L$ chains under periodic boundary conditions. Inset: enlarged
figure around the lower level crossing. (b) Finite-size scaling analysis
of the level-crossing points for the upper bound of the BN phase.  (c)
Ground-state energy per site as a function of $W/t$ for the 1D model at
$U/t=2$ in the thermodynamic limit, which is obtained using DMRG data
with $L=48$-$240$ chains under the open boundary conditions.}
\label{figDMRG_1D}
\end{center}
\end{figure}

Therefore, to confirm the BN and FM states and to explore the phase
boundaries, we calculate the ground-state energy by the numerical
methods.  In Fig.~\ref{figDMRG_1D}(a) the numerical results of the
ground-state energy at $U=2t$ are plotted as a function of $W/t$, where
the periodic boundary conditions are applied. We obtain numerically the
BN ground-state energy $\varepsilon_0=U/2\equiv\varepsilon_0({\rm BN})$
for $W \ge U/2$ and the FM ground-state energy
$\varepsilon_0=2W\equiv\varepsilon_0({\rm FM})$ for $W \le U/2$. Thus,
the BN-FM phase boundary coincides the analytical result $W=U/2$. The
ground-state energy deviates from $\varepsilon_0=\varepsilon_0({\rm
BN})$ at some larger $W/t$ ($\equiv W_c/t$), which corresponds to the
upper bound of the BN phase and is detected as a level crossing in the
present finite-$L$ calculations. As seen in Fig.~\ref{figDMRG_1D}(a),
the level-crossing point depends on the system length because the BN
state is more overstabilized in smaller-$L$ systems under the periodic
boundary conditions. Accordingly, the level-crossing point is shifted to
lower $W/t$ with increasing the system length $L$.  We perform a
finite-size scaling of the level-crossing point using $L=26$-$50$
periodic systems in Fig.~\ref{figDMRG_1D}(b). Although the data points
oscillate and a fine fitting is not easy, the least-square linear
fitting gives $W_c/t=6.37$ in the thermodynamic limit. This may mean the
upper bound of the BN phase is fairly extended to $W/t=6.37$ in
comparison to the analytical value $W/t=3$ in
Fig.~\ref{phase_diagrams_1D}.

The above overstabilization of the BN state can be avoided if we apply
the open boundary conditions. It enables us to pick up the {\it real}
ground state and to calculate the energy more definitely for a given
$L$.  The extrapolated ground-state energy to the thermodynamic limit,
using $L=48$-$240$ open systems, is plotted in Fig.~\ref{figDMRG_1D}(c).
We find that the ground-state energy begins to deviate from
$\varepsilon_0=\varepsilon_0({\rm BN})$ at $W/t=6.40$. This value agrees
very well with that obtained with the periodic systems ($W/t=6.37$).

\section{Kagom\'e lattice}\label{sec:2D}

\begin{figure}[t]
 \input{fig4.tex}
 \caption{Three bases for the unit trimer of the Kagom\'e lattice.}
\label{fig:threeB}
\end{figure}
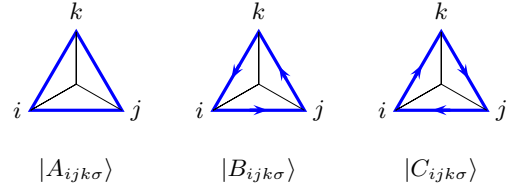

We consider the generalized Hubbard model on the Kagom\'e lattice at
$1/3$-filling with zero-magnetic field. In order to obtain an exact
ground state, we need to include three site terms ($X'$, $W'$ terms).
The Hamiltonian is given by ${\cal H}=\sum_{\langle
ijk\rangle\sigma}h_{ijk\sigma}$, where the summation $\langle
ijk\rangle$ is taken in each unit trimer as shown in
Fig.~\ref{fig:lattices},
\begin{align}
 \lefteqn{h_{ijk\sigma}
 =h_{ij\sigma}+h_{jk\sigma}+h_{ki\sigma}}
  \nonumber\\
 &
  +W'(T_{ij\sigma}T_{jk\bar{\sigma}}+T_{jk\sigma}T_{ki\bar{\sigma}}
  +T_{ki\sigma}T_{ij\bar{\sigma}})\nonumber\\
 &
 +X'(T_{ij\sigma}n_{k\bar{\sigma}}+T_{jk\sigma}n_{i\bar{\sigma}}
 +T_{ki\sigma}n_{j\bar{\sigma}}),
 \label{local_trim_Ham}
\end{align}
where $h_{ij\sigma}$ is the local bond Hamiltonian
(\ref{local_bond_Ham}) with $z=2$. $\bar{\sigma}$ denotes the opposite
spin of $\sigma$.
Now we define the following one-electron plaquette operators (see
Fig.~\ref{fig:threeB}),
\begin{align}
 A_{ijk\sigma}^{\dag}
  \equiv&{\textstyle\frac{1}{\sqrt{3}}}
  (c_{i\sigma}^{\dag}+c_{j\sigma}^{\dag}+c_{k\sigma}^{\dag}),\\
 B_{ijk\sigma}^{\dag}
  \equiv&{\textstyle\frac{1}{\sqrt{3}}}
  (c_{i\sigma}^{\dag}+\omega c_{j\sigma}^{\dag}
  +\omega^2 c_{k\sigma}^{\dag}),\\
 C_{ijk\sigma}^{\dag}
  \equiv&{\textstyle\frac{1}{\sqrt{3}}}
  (c_{i\sigma}^{\dag}+\omega^2c_{j\sigma}^{\dag}
  +\omega c_{k\sigma}^{\dag}),
\end{align}
where $\omega=\e^{\i 2\pi/3}$.  These operators on the same plaquette
satisfy the anticommutation relations:
\begin{displaymath}
 \{A_{ijk\sigma},A_{ijk\sigma'}^{\dag}\}
 =\{B_{ijk\sigma},B_{ijk\sigma'}^{\dag}\}
 =\{C_{ijk\sigma},C_{ijk\sigma'}^{\dag}\}
 =\delta_{\sigma\sigma'},
\end{displaymath}
and otherwise$=0$.  Note that $A_{ijk\sigma}^{\dag}|0\rangle$,
$B_{ijk\sigma}^{\dag}|0\rangle$, and $C_{ijk\sigma}^{\dag}|0\rangle$ are
chosen as eigen states of density, hopping, and current operators,
\begin{align}
 N_{ijk\sigma}\equiv&n_{i\sigma}+n_{j\sigma}+n_{k\sigma},\\
 T_{ijk\sigma}\equiv&T_{ij\sigma}+T_{jk\sigma}+T_{ki\sigma},\\
 J_{ijk\sigma}\equiv&J_{ij\sigma}+J_{jk\sigma}+J_{ki\sigma},\\
 J_{ij\sigma}\equiv&
  \i(c_{i\sigma}^{\dag}c_{j\sigma}^{\mathstrut}-\mbox{H.c.}).
\end{align}
The density operators in terms of the plaquette operators are
\begin{align}
 n_{A\sigma}=&\frac{1}{3}
  \left(N_{ijk\sigma}+T_{ijk\sigma}\right),\\
 n_{B\sigma}=&\frac{1}{6}
 (2N_{ijk\sigma}-T_{ijk\sigma}-\sqrt{3}J_{ijk\sigma}),\\
 n_{C\sigma}=&\frac{1}{6}
 (2N_{ijk\sigma}-T_{ijk\sigma}+\sqrt{3}J_{ijk\sigma}).
\end{align}
Since we restrict our attention only on the three sites $i,j,k$ in a
triangle, we drop these indices from the operators defined above.

\subsection{Plaquette-N\'eel state at $1/3$-filling}

Using these relations, the Hamiltonian with the exact ground state is
given by the plaquette operators.  We consider the following plaquette
state at $1/3$-filling,
\begin{equation}
 \ket{\Psi_{\sigma}}\equiv
 \prod_{\braket{ijk}\in\bigtriangleup} A_{ijk\sigma}^{\dag}
  \prod_{\braket{i'j'k'}\in\bigtriangledown}
  A_{i'j'k'\bar{\sigma}}^{\dag}\ket{0},
  \label{Kagome_13_state}
\end{equation}
where $\braket{ijk}$ ($\braket{i'j'k'}$) is taken for all triangles of
the Kagom\'e lattice with up (down) direction. As an extention of the BN
state, we call this state ``plaquette N\'eel'' (PN) state.

\begin{figure}[t]
 \input{fig5.tex}
 \caption{Phase diagrams of the generalized Hubbard model on the
 Kagom\'e lattice, in the $U/|t|$-$W/|t|$ parameter space for (a) $t>0$
 at $1/3$-filling~\cite{Nakamura-I_2004} and (b) $t<0$ at $2/3$-filling,
 respectively.  The shaded regions labeled by PN denote the plaquette
 N\'eel state.}  \label{fig:PNpd}
\end{figure}
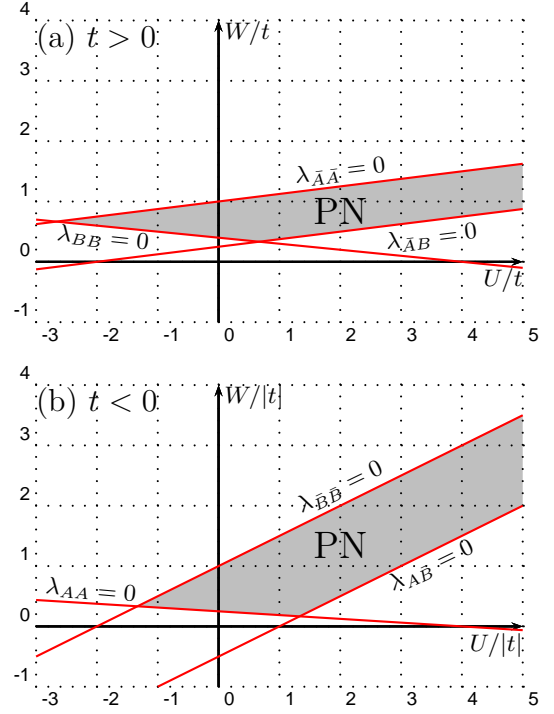

In order to make (\ref{Kagome_13_state}) the ground state, the local
Hamiltonian for this state is constructed as
\begin{align}
\lefteqn{
 h_{ijk}-\varepsilon_0=
  \lambda_{\bar{A}\bar{A}} (1-n_{A\uparrow}) (1-n_{A\downarrow})}
  \label{decomposed_Hamiltonian_2D}\\
 &
  +\lambda_{BB}n_{B\uparrow}n_{B\downarrow}
  +\lambda_{CC}n_{C\uparrow}n_{C\downarrow}\nonumber\\
 &+\lambda_{\bar{A}B}\left\{
      (1-n_{A\uparrow})n_{B\downarrow}+
      n_{B\uparrow}(1-n_{A\downarrow})\right\}\nonumber\\
 &+\lambda_{\bar{A}C}\left\{
      (1-n_{A\uparrow})n_{C\downarrow}+
      n_{C\uparrow}(1-n_{A\downarrow})\right\}\nonumber\\
 &+\lambda_{BC}
  \left\{
   n_{B\uparrow}n_{C\downarrow}+n_{C\uparrow}n_{B\downarrow}
  \right\}
 \nonumber\\
 =&\lambda_{\bar{A}\bar{A}}
  \label{decomposed_Hamiltonian_2D.2}\\
 &+\sum_{\sigma}
  \left\{
   -\lambda_{\bar{A}\bar{A}}n_{A\sigma}
   +\lambda_{\bar{A}B}n_{B\sigma}
   +\lambda_{\bar{A}C}n_{C\sigma}
  \right\}\nonumber\\
 &+\lambda_{\bar{A}\bar{A}} n_{A\uparrow}n_{A\downarrow}
  +\lambda_{BB}n_{B\uparrow}n_{B\downarrow}
  +\lambda_{CC}n_{C\uparrow}n_{C\downarrow}\nonumber\\
 &-\lambda_{\bar{A}B}
  (n_{A\uparrow}n_{B\downarrow}+n_{B\uparrow}n_{A\downarrow})
 \nonumber\\
 &
  -\lambda_{\bar{A}C}
  (n_{A\uparrow}n_{C\downarrow}+n_{C\uparrow}n_{A\downarrow})
 \nonumber\\
 &
  +\lambda_{BC}
  (n_{B\uparrow}n_{C\downarrow}+n_{C\uparrow}n_{B\downarrow}),
 \nonumber
\end{align}
with positive $\lambda_{\mu\nu}$. Here we consider the case that
$\lambda_{BB}=\lambda_{CC}$ and $\lambda_{\bar{A}B}=\lambda_{\bar{A}C}$,
assuming the time-reversal symmetry of the Hamiltonian. Then we have
\begin{align}
 \lefteqn{
 h_{ijk}-\varepsilon_0
 =\frac{1}{3}(\lambda_{\bar{A}\bar{A}}+\lambda_{\bar{A}B}) h_{t}}
  \label{decomposed_Hamiltonian_2D.3}\\
 &
  +\frac{1}{9}\left(\lambda_{\bar{A}\bar{A}}
	       -4\lambda_{\bar{A}B}+4\lambda_{BB}\right)
  (2h_{U}+h_{V_{\perp}})\nonumber\\
 &
  +\frac{1}{9}\left(\lambda_{\bar{A}\bar{A}}
	       +2\lambda_{\bar{A}B}+\lambda_{BB}\right)
  (h_{V_{\parallel}}+h_{W}+h_{W'})\nonumber\\
 &
  +\frac{1}{9}\left(\lambda_{\bar{A}\bar{A}}
       -\lambda_{\bar{A}B}-2\lambda_{BB}\right)(h_{X}+h_{X'})\nonumber\\
 &
 +\frac{1}{9}\left(-4\lambda_{\bar{A}\bar{A}}+4\lambda_{\bar{A}B}
 -\lambda_{BB}\right)\sum_{\sigma}N_{ijk\sigma}
 +\lambda_{\bar{A}\bar{A}},\nonumber
\end{align}
where $h_t$, $h_U$, $\cdots$, $h_{X'}$ are defined in
Appendix~\ref{details}.  For $1/3$-filling, the density operator and the
number of the triangles $N_{\rm tr}$ is related as
$\sum_{\braket{ijk},\sigma}N_{ijk\sigma}=2N_{\rm tr}$, and the number of
lattice sites is $L=3N_{\rm tr}/2$, so that the ground-state energy per
site is identified as
\begin{equation}
 \varepsilon_0=\frac{1}{9}
  \left(\lambda_{\bar{A}\bar{A}}+8\lambda_{\bar{A}B}-2\lambda_{BB}\right).
\end{equation}
The coefficients of projection operators are related to the parameters
as
\begin{align}
 \left[\begin{array}{c}
  \lambda_{\bar{A}\bar{A}}\\
	\lambda_{\bar{A}B}\\
	\lambda_{BB}
       \end{array}\right]
 =&\left[\begin{array}{ccc}1&4&4\\ -1&2&-1\\ 1&1&-2
	\end{array}\right]
       \left[\begin{array}{c}U/2\\W\\X\end{array}\right].
 \label{matrix_relation_PN}
\end{align}
Using the condition of the hopping in
Eq.~(\ref{decomposed_Hamiltonian_2D.3}), we have
\begin{align}
 \lambda_{\bar{A}\bar{A}}=&\frac{U}{2}-4W+4t,\\
 \lambda_{\bar{A}B}=&-\frac{U}{2}+4W-t,\\
 \lambda_{BB}=&\frac{U}{2}+5W-2t.
\end{align}
Since all these coefficients should be positive, the condition of the
exact PN ground state is given as follows,
\begin{gather}
 W\leq\frac{U}{8}+t,\quad
 W\geq\frac{U}{8}+\frac{t}{4},\quad
 W\geq -\frac{U}{10}+\frac{2t}{5},\nonumber\\
 V_{\perp}=\frac{U}{2},\quad V_{\parallel}=W=W',\quad
 X=X'=t-2W,\nonumber
\end{gather}
and the ground state energy per site is
\begin{equation}
 \varepsilon_0=\frac{1}{3}(U-4W).
\label{PNenergy_n13}
\end{equation}
The phase diagram for the exact PN state is surrounded by three lines
given by $\lambda_{\bar{A}\bar{A}}>0$, $\lambda_{\bar{A}B}>0$, and
$\lambda_{BB}>0$, as shown in Fig~\ref{fig:PNpd}(a).

\begin{figure}
\begin{center}
\includegraphics[width=7cm]{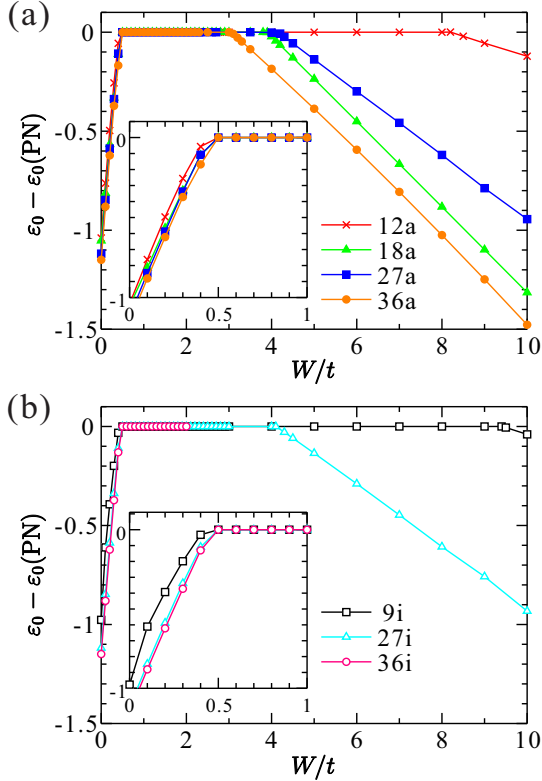}
\caption{(a) Ground-state energy per site as a function of $W/t$
for the $1/3$-filling Kagom\'e model with $U/t=2$ and $t>0$, obtained
by the numerical methods using (a) anisotropic and (b) isotropic clusters
(see Appendix B). The energy of the PN state [Eq.(\ref{PNenergy_n13})]
is subtracted. Insets: enlarged figure around the lower level crossing.
}
\label{figDMRG_kagome_n13}
\end{center}
\end{figure}

In Fig.~\ref{figDMRG_kagome_n13}, the numerical results of the
ground-state energy for the $1/3$-filling Kagom\'e model at $U/t=2$ with
$t>0$ are plotted as a function of $W/t$. The energy of the PN state
$\varepsilon_0=(U-4W)/3\equiv\varepsilon_0({\rm PN})$ is subtracted so
that a region with $\varepsilon_0-\varepsilon_0({\rm PN})=0$ corresponds
to the PN phase.  As seen in Figs.~\ref{figDMRG_kagome_n13}(a) and (b),
a robust range with $\varepsilon_0-\varepsilon_0({\rm PN})=0$ exists for
all the used clusters. We find a deviation from
$\varepsilon_0-\varepsilon_0({\rm PN})=0$ below a level-crossing point
$W/t=0.5$, which is independent of the cluster shape and size [see
insets of Figs.~\ref{figDMRG_kagome_n13}(a) and (b)]. This $W/t$ value
agrees perfectly with the analytical result of the lower bound of the PN
phase, given by $\lambda_{\bar{A}B}=0$. Whereas, the level-crossing
point related to the upper bound depends on the cluster.  Like in the 1D
BN state, the PN state would be overstabilized with small
clusters. However, the data are not sufficient to perform a finite-size
scaling analysis and it remains as a future work.

\subsection{Plaquette-N\'eel state at $2/3$-filling}

We consider the following plaquette N\'eel state at $2/3$-filling given
by
\begin{equation}
 \ket{\Psi_{\sigma}}\equiv
 \prod_{\braket{ijk}\in\bigtriangleup}
 C_{ijk\sigma}^{\dag}
 B_{ijk\sigma}^{\dag}
  \prod_{\braket{i'j'k'}\in\bigtriangledown}
  C_{i'j'k'\bar{\sigma}}^{\dag}
  B_{i'j'k'\bar{\sigma}}^{\dag}
  \ket{0},
  \label{Kagome_23_state}
\end{equation}
where $\braket{ijk}$ ($\braket{i'j'k'}$) is taken for all triangles of
the Kagom\'e lattice with up (down) direction, and
\begin{equation}
  C_{ijk\sigma}^{\dag}B_{ijk\sigma}^{\dag}
  ={\textstyle\frac{\i}{\sqrt{3}}}
  (c_{i\sigma}^{\dag}c_{j\sigma}^{\dag}
  +c_{j\sigma}^{\dag}c_{k\sigma}^{\dag}
  +c_{k\sigma}^{\dag}c_{i\sigma}^{\dag}).
\end{equation}
The Hamiltonian for this state is constructed as
\begin{align}
\lefteqn{
 h_{ijk}-\varepsilon_0=
  \lambda_{AA} n_{A\uparrow} n_{A\downarrow}}\\
 &
  +\lambda_{\bar{B}\bar{B}}(1- n_{B\uparrow})(1- n_{B\downarrow})
  +\lambda_{\bar{C}\bar{C}}(1- n_{C\uparrow})(1- n_{C\downarrow})\nonumber\\
 &+\lambda_{A\bar{B}}\left\{
      n_{A\uparrow}(1-n_{B\downarrow})+
      (1-n_{B\uparrow})n_{A\downarrow}\right\}\nonumber\\
 &+\lambda_{A\bar{C}}\left\{
      n_{A\uparrow}(1-n_{C\downarrow})+
      (1-n_{C\uparrow})n_{A\downarrow}\right\}\nonumber\\
 &+\lambda_{\bar{B}\bar{C}}
  \left\{
   (1- n_{B\uparrow})(1- n_{C\downarrow})
   +(1- n_{C\uparrow})(1- n_{B\downarrow})
  \right\}\nonumber\\
 =&\lambda_{\bar{B}\bar{B}}+\lambda_{\bar{C}\bar{C}}
 +2\lambda_{\bar{B}\bar{C}}\\
 &
  +\sum_{\sigma}
  \{
   (\lambda_{A\bar{B}}+\lambda_{A\bar{C}})n_{A\sigma}
   -(\lambda_{\bar{B}\bar{B}}+\lambda_{\bar{B}\bar{C}})n_{B\sigma}\nonumber\\
 &
   -(\lambda_{\bar{C}\bar{C}}+\lambda_{\bar{B}\bar{C}})n_{C\sigma}
   \}\nonumber\\
 &+\lambda_{AA} n_{A\uparrow}n_{A\downarrow}
  +\lambda_{\bar{B}\bar{B}}n_{B\uparrow}n_{B\downarrow}
  +\lambda_{\bar{C}\bar{C}}n_{C\uparrow}n_{C\downarrow}\nonumber\nonumber\\
 &-\lambda_{A\bar{B}}
  (n_{A\uparrow}n_{B\downarrow}+n_{B\uparrow}n_{A\downarrow})\nonumber\\
 &
  -\lambda_{A\bar{C}}
  (n_{A\uparrow}n_{C\downarrow}+n_{C\uparrow}n_{A\downarrow})\nonumber\\
 &
  +\lambda_{\bar{B}\bar{C}}
  (n_{B\uparrow}n_{C\downarrow}+n_{C\uparrow}n_{B\downarrow}).
 \nonumber
\end{align}
For $\lambda_{\bar{B}\bar{B}}=\lambda_{\bar{B}\bar{C}}
=\lambda_{\bar{C}\bar{C}}$, and $\lambda_{A\bar{B}}=\lambda_{A\bar{C}}$,
assuming the time-reversal symmetry of the Hamiltonian, we have
\begin{align}
 \lefteqn{
 h_{ijk}-\varepsilon_0
 =\frac{2}{3}(-\lambda_{A\bar{B}}-\lambda_{\bar{B}\bar{B}}) h_{t}}
  \label{decomposed_Hamiltonian_2D.5}\\
 &
  +\frac{1}{9}\left(\lambda_{AA}-4\lambda_{A\bar{B}}
 +4\lambda_{\bar{B}\bar{B}}\right)(2h_{U}+h_{V_{\perp}})\nonumber\\
 &
 +\frac{1}{9}\left(\lambda_{AA}+2\lambda_{A\bar{B}}
 +\lambda_{\bar{B}\bar{B}}\right)
  (h_{V_{\parallel}}+h_{W}+h_{W'})\nonumber\\
 &
 +\frac{1}{9}\left(\lambda_{AA}-\lambda_{A\bar{B}}
 -2\lambda_{\bar{B}\bar{B}}\right)(h_{X}+h_{X'})\nonumber\\
 &
 +\frac{1}{9}\left(-
\lambda_{AA}+4\lambda_{A\bar{B}}
 -13\lambda_{\bar{B}\bar{B}}\right)
 \sum_{\sigma}N_{ijk\sigma}
 +4\lambda_{\bar{B}\bar{B}}.\nonumber
\end{align}
For $2/3$-filling, the density operator and the number of the triangles
$N_{\rm tr}$ is related as
$\sum_{\braket{ijk},\sigma}N_{ijk\sigma}=4N_{\rm tr}$, and the number of
lattice sites is $L=3N_{\rm tr}/2$, so that the ground-state energy per
site is identified as
\begin{equation}
 \varepsilon_0=
  \frac{8}{27}(\lambda_{AA}-4\lambda_{A\bar{B}}+4\lambda_{\bar{B}\bar{B}}).
\end{equation}
Since the relation between $(\lambda_{AA}, \lambda_{A\bar{B}},
\lambda_{\bar{B}\bar{B}})$ and $(U, W, X)$ is given by the same matrix
as that of (\ref{matrix_relation_PN}), we identify the coefficients of
the projection operators, using the condition for the hopping in
Eq.~(\ref{decomposed_Hamiltonian_2D.5}), as
\begin{align}
 \lambda_{AA}=&\frac{U}{2}+8W+2t,\\
 \lambda_{A\bar{B}}=&-\frac{U}{2}+W-\frac{t}{2},\\
 \lambda_{\bar{B}\bar{B}}=&\frac{U}{2}-W-t.
\end{align}
Thus the condition of the exact PN ground state is given as follows,
\begin{gather}
 W\geq-\frac{U}{16}-\frac{t}{4},\quad
 W\geq \frac{U}{2}+\frac{t}{2},\quad
 W\leq \frac{U}{2}-t,\nonumber\\
 V_{\perp}=\frac{U}{2},\quad V_{\parallel}=W=W',\quad
 X=X'=\frac{t}{2}+W.
\end{gather}
The ground state energy per site is
\begin{equation}
 \varepsilon_0=\frac{4}{3}U.
\end{equation}
The phase diagram for the exact plaquette N\'eel state is surrounded by
three boundaries given by $\lambda_{AA}>0$, $\lambda_{A\bar{B}}>0$, and
$\lambda_{\bar{B}\bar{B}}>0$, as shown in Fig~\ref{fig:PNpd}(b).

\begin{figure}
\begin{center}
\includegraphics[width=7cm]{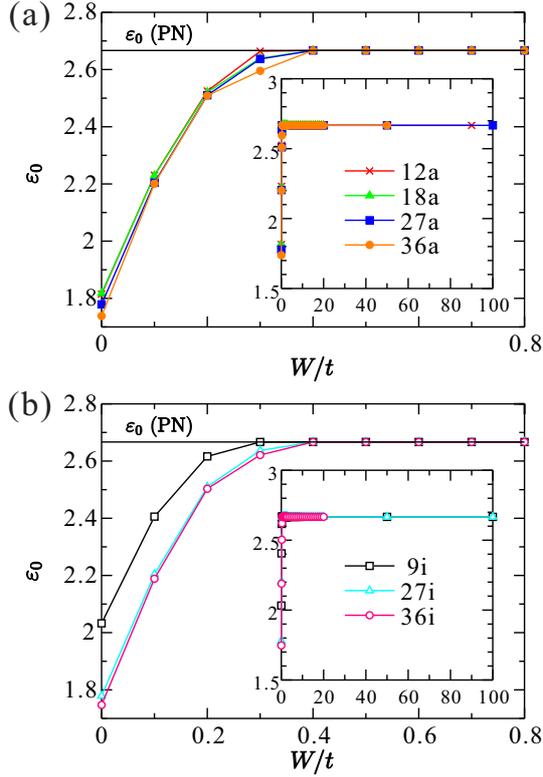}
\caption{(a) Ground-state energy per site as a function of $W/|t|$
for the $2/3$-filling Kagom\'e model with $U/|t|=2$ and $t<0$,
obtained by the numerical methods using (a) anisotropic and (b) isotropic
clusters (see Appendix B).
Insets: similar figures for a wider range of $W/|t|$.
}
\label{figDMRG_kagome_n23}
\end{center}
\end{figure}

The numerical results of the ground-state energy for the $2/3$-filling
Kagom\'e model at $U/|t|=2$ with $t<0$ are plotted as a function of
$W/|t|$ in Fig.~\ref{figDMRG_kagome_n23}. We find that the system has
the PN state energy $\varepsilon_0({\rm PN})=4U/3$ in a wide range of
$W/|t|$.  The energy deviation from $\varepsilon_0=\varepsilon_0({\rm
PN})$, indicating a transition to another phase, is clearly
seen. Although the level crossing is not very sharp, we can
approximately estimate the transition point $W/|t|\sim0.4$ for all the
used clusters. This value is close but subtly smaller than the
analytical result of the lower bound of the PN phase
$W/|t|=U/(2|t|)-1/2=0.5$ given by $\lambda_{A\bar{B}}=0$.  Let us then
turn to the upper bound of the PN phase. It may be more puzzling.
Differently from the case of $1/3$-filling Kagom\'e lattice with $t>0$,
the PN state seems to maintain as the ground state up to $W/|t|=100$ in
the present calculations with periodic clusters. To resolve this issue,
further calculations are required.

\subsection{Ferromagnetism at $1/2$-filling}

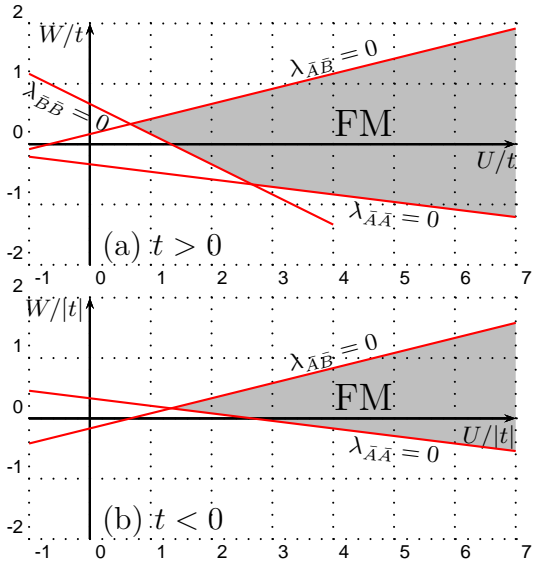
\begin{figure}[t]
 \input{fig8.tex}
 \caption{
 Phase diagrams of the generalized Hubbard model on the Kagom\'e
 lattice, in the $U/|t|$-$W/|t|$ parameter space with
 $V_{\parallel}=W=W'$, $V_{\perp}=U/2$, $X=X'=t/3$.  (a) and (b)
 correspond to the case of $t>0$ and $t<0$, respectively. The shaded
 regions labeled by FM denote the exact ferromagnetic ground state.
 }\label{fig:FMpd}
\end{figure}

We consider a ferromagnetic (FM) state at half-filling where each
triangle is occupied by three particles with the same spin,
\begin{equation}
 \ket{\Psi_{\sigma}}
  \equiv
  \prod_{\braket{ijk}\in\bigtriangleup}
  C_{ijk\sigma}^{\dag}B_{ijk\sigma}^{\dag}A_{ijk\sigma}^{\dag}\ket{0}.
\end{equation}
where
\begin{equation}
 C_{ijk\sigma}^{\dag}B_{ijk\sigma}^{\dag}A_{ijk\sigma}^{\dag}
 =\i c_{i\sigma}^{\dag}c_{j\sigma}^{\dag}c_{k\sigma}^{\dag}.
\end{equation}
The Hamiltonian for this state is constructed as
\begin{align}
\lefteqn{
 h_{ijk}-\varepsilon_0=
  \lambda_{\bar{A}\bar{A}} (1-n_{A\uparrow}) (1-n_{A\downarrow})}\\
 &+\lambda_{\bar{B}\bar{B}}(1- n_{B\uparrow})(1- n_{B\downarrow})\nonumber\\
 &
  +\lambda_{\bar{C}\bar{C}}(1- n_{C\uparrow})(1- n_{C\downarrow})\nonumber\\
 &+\lambda_{\bar{A}\bar{B}}\left\{
      (1-n_{A\uparrow})(1-n_{B\downarrow})+
      (1-n_{B\uparrow})(1-n_{A\downarrow})\right\}\nonumber\\
 &+\lambda_{\bar{A}\bar{C}}\left\{
      (1-n_{A\uparrow})(1-n_{C\downarrow})+
      (1-n_{C\uparrow})(1-n_{A\downarrow})\right\}\nonumber\\
 &+\lambda_{\bar{B}\bar{C}}
  \left\{
   (1-n_{B\uparrow})(1-n_{C\downarrow})
   +(1-n_{C\uparrow})(1-n_{B\downarrow})
  \right\}\nonumber\\
 =&
 \lambda_{\bar{A}\bar{A}}
 +\lambda_{\bar{B}\bar{B}}
 +\lambda_{\bar{C}\bar{C}}
 +2(\lambda_{\bar{A}\bar{B}}+\lambda_{\bar{B}\bar{C}}
 +\lambda_{\bar{A}\bar{C}})\\
 &+\sum_{\sigma}
  \{
   -(\lambda_{\bar{A}\bar{A}}+\lambda_{\bar{A}\bar{B}}
   +\lambda_{\bar{A}\bar{C}})n_{A\sigma}\nonumber\\
 &
   -(\lambda_{\bar{B}\bar{B}}+\lambda_{\bar{A}\bar{B}}
   +\lambda_{\bar{B}\bar{C}})n_{B\sigma}\nonumber\\
 &
   -(\lambda_{\bar{C}\bar{C}}+\lambda_{\bar{A}\bar{C}}
   +\lambda_{\bar{B}\bar{C}})n_{C\sigma}
   \}\nonumber\\
 &+\lambda_{\bar{A}\bar{A}} n_{A\uparrow}n_{A\downarrow}
  +\lambda_{\bar{B}\bar{B}}n_{B\uparrow}n_{B\downarrow}
  +\lambda_{\bar{C}\bar{C}}n_{C\uparrow}n_{C\downarrow}\nonumber\\
 &+\lambda_{\bar{A}\bar{B}}
  (n_{A\uparrow}n_{B\downarrow}+n_{B\uparrow}n_{A\downarrow})\nonumber\\
 &
  +\lambda_{\bar{A}\bar{C}}
  (n_{A\uparrow}n_{C\downarrow}+n_{C\uparrow}n_{A\downarrow})\nonumber\\
 &
  +\lambda_{\bar{B}\bar{C}}
  (n_{B\uparrow}n_{C\downarrow}+n_{C\uparrow}n_{B\downarrow}).
 \nonumber
\end{align}
Under the time-reversal symmetry
$\lambda_{\bar{B}\bar{B}}=\lambda_{\bar{B}\bar{C}}=\lambda_{\bar{C}\bar{C}}$,
and $\lambda_{\bar{A}\bar{B}}=\lambda_{\bar{A}\bar{C}}$, we have
\begin{align}
 \lefteqn{
 h_{ijk}-\varepsilon_0
 =\frac{1}{3}(\lambda_{\bar{A}\bar{A}}+\lambda_{\bar{A}\bar{B}}
  -2\lambda_{\bar{B}\bar{B}})h_{t}}\label{Ham_ferro.3}\\
 &
  +\frac{1}{9}(\lambda_{\bar{A}\bar{A}}+4\lambda_{\bar{A}\bar{B}}
  +4\lambda_{\bar{B}\bar{B}})(2h_{U}+h_{V_{\perp}})\nonumber\\
 &
 +\frac{1}{9}(\lambda_{\bar{A}\bar{A}}-2\lambda_{\bar{A}\bar{B}}
  +\lambda_{\bar{B}\bar{B}})(h_{V_{\parallel}}+h_{W}+h_{W'})\nonumber\\
 &
 +\frac{1}{9}\left(\lambda_{\bar{A}\bar{A}}+\lambda_{\bar{A}\bar{B}}
 -2\lambda_{\bar{B}\bar{B}}\right)(h_{X}+h_{X'})\nonumber\\
 &
 -\frac{1}{9}\left(4\lambda_{\bar{A}\bar{A}}+10\lambda_{\bar{A}\bar{B}}
 +13\lambda_{\bar{B}\bar{B}}\right)\sum_{\sigma}N_{ijk\sigma}\nonumber\\
 &+\lambda_{\bar{A}\bar{A}}+4\lambda_{\bar{A}\bar{B}}
 +4\lambda_{\bar{B}\bar{B}}.
 \nonumber
\end{align}
At half-filling, the number of the triangle $N_{\rm tr}$ is related as
 $\sum_{\braket{ijk},\sigma}N_{ijk\sigma}=3N_{\rm tr}$, and the number
 of lattice sites $L=3N_{\rm tr}/2$, so that the ground-state energy per
 site is identified as
\begin{equation}
 \varepsilon_0=
  \frac{2}{9}(\lambda_{\bar{A}\bar{A}}-2\lambda_{A\bar{B}}
  +\lambda_{\bar{B}\bar{B}}).
\end{equation}
The parameters are related as
\begin{align}
 \left[\begin{array}{c}
  \lambda_{\bar{A}\bar{A}}\\
	\lambda_{\bar{A}\bar{B}}\\
	\lambda_{\bar{B}\bar{B}}
       \end{array}\right]
 =&\left[\begin{array}{ccc}1&4&4\\ 1&-2&1\\ 1&1&-2
	\end{array}\right]
       \left[\begin{array}{c}U/2\\W\\X\end{array}\right].
\end{align}
Then we have
\begin{align}
 \lambda_{\bar{A}\bar{A}}=&\frac{U}{2}+4W+\frac{4}{3}t,\\
 \lambda_{\bar{A}\bar{B}}=&\frac{U}{2}-2W+\frac{t}{3},\\
 \lambda_{\bar{B}\bar{B}}=&\frac{U}{2}+W-\frac{2}{3}t.
\end{align}
Thus the
condition of the exact ferromagnetic ground state is given as follows,
\begin{gather}
 W\geq-\frac{U}{8}-\frac{t}{3},\quad
 W\leq \frac{U}{4}+\frac{t}{6},\quad
 W\geq-\frac{U}{2}+\frac{2}{3}t,\nonumber\\
 V_{\perp}=\frac{U}{2},\quad
 V_{\parallel}=W=W',\quad
 X=X'=\frac{t}{3}.
\end{gather}
The ground state energy per site is
\begin{equation}
 \varepsilon_0=2W.
\end{equation}
This is consistent with the fact that in the fully ferromagnetic state,
only the $V_{\parallel}$ term contribute to the energy.  The condition
of the hopping in Eq.~(\ref{Ham_ferro.3}) means that $t$ may take both
positive and negative values.  As shown in Fig~\ref{fig:FMpd}, (a) for
positive $t$, the exact ferromagnetic ground state is surrounded by
three lines, while (b) for the negative $t$, the lines become two.

\begin{figure}
\begin{center}
\includegraphics[width=7cm]{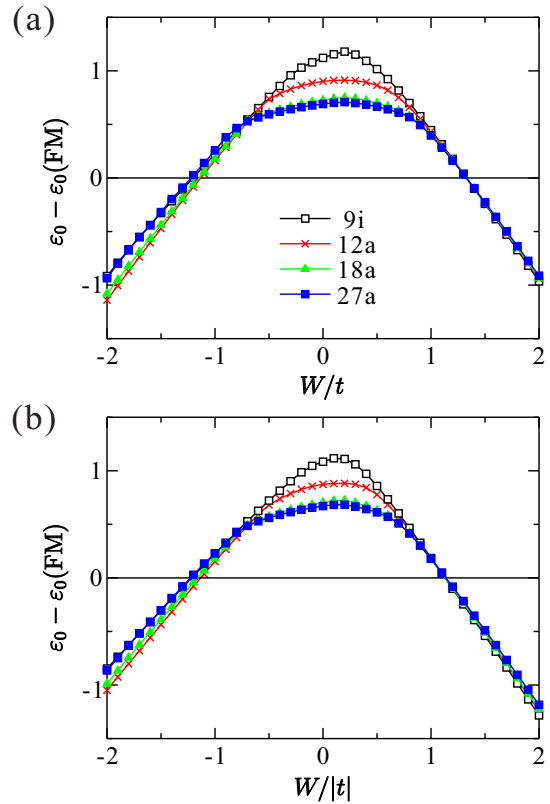}
\caption{Numerical results of the ground-state energy per site as a function
of $W/|t|$ for the $1/2$-filling Kagom\'e model at $U/|t|=3$ with (a) $t>0$
and (b) $t<0$, where the number of spin-up and spin-down electrons are
kept as $N_\uparrow=N_\downarrow$ or $N_\uparrow=N_\downarrow+1$.
The energy of the FM state $\varepsilon_0({\rm FM})=2W$ is subtracted.
}
\label{figDMRG_kagome_n12}
\end{center}
\end{figure}

In Fig.~\ref{figDMRG_kagome_n12} the numerical results of the ground-state
energy for the $1/2$-filling Kagom\'e model at $U/|t|=3$ are plotted as
a function of $W/|t|$, where the numbers of spin-up and spin-down
electrons are kept to be as close as possible, namely,
$N_\uparrow-N_\downarrow=0$ and $|N_\uparrow-N_\downarrow|=1$ for even-
and odd-site clusters, respectively. Since the ground-state energy of
the FM state $\varepsilon_0=2W\equiv\varepsilon_0({\rm FM})$ is
subtracted in Fig.~\ref{figDMRG_kagome_n12}, the FM phase is indicated
by a region having positive value of the numerical energy
$\varepsilon_0-\varepsilon_0({\rm FM})>0$.  The finite-size effect seems
to be much smaller than that in the PN state. For the both positive and
negative $t$ values, the FM phase appears at $-1 \lesssim W/|t|\lesssim 1$,
though the region for $t<0$ may be slightly narrower than that for $t>0$.
The FM phase would be comparatively more extended than the
analytical results shown in Fig~\ref{fig:FMpd}.

\section{Entanglement entropy} \label{sec:EE}

In this section we consider the entanglement entropy
(EE)\cite{Horodecki-H-H-H} of the system discussed above.  When we
divide the normalized wave function of the system into two regions A and
B as
\begin{equation}
 \ket{\Psi}=\sum_{nm}\Lambda_{nm}
  \ket{\Psi_n^{\rm A}}\otimes\ket{\Psi_m^{\rm B}},
\end{equation}
the EE is given by
\begin{equation}
 S^{\rm A} =-\mathrm{Tr}_{\rm A}
  \left[ \hat{\rho}_{\rm A}\log\hat{\rho}_{\rm A}\right],
\end{equation}
with the reduced density matrix
\begin{equation}
 \hat{\rho}_{\rm A}=\sum_{nm}(\Lambda \Lambda^T)_{nm}
  \ket{\Psi^{\rm A}_n}\bra{\Psi^{\rm A}_m},
  \label{RDM}
\end{equation}
where $\Lambda^T$ is the transposed matrix of $\Lambda$.

\begin{figure}[t]
 \input{fig10.tex}
\caption{Patterns to cut the systems into two regions A and B to
calculate the entanglement entropy (EE) for (a) the 1D chain and (b) the
Kagom\'e lattice, respectively.}\label{fig:dividing}
\end{figure}
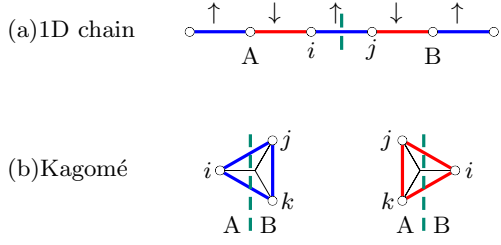

For the BN state in 1D, $\ket{\Psi_n^{\rm A}}$ and $\ket{\Psi_n^{\rm
B}}$ (see Fig.~\ref{fig:dividing}(a)) are given as
\begin{align}
 \ket{\Psi_1^{\rm A}}
 =&X_{\rm A}^{\dag} c_{i\sigma}^\dag \ket{0}_{\rm A},\\
 \ket{\Psi_2^{\rm A}}
 =&X_{\rm A}^{\dag}\ket{0}_{\rm A},\\
 \ket{\Psi_1^{\rm B}}
 =&c_{j\sigma}^\dag X_{\rm B}^{\dag} \ket{0}_{\rm B},\\
 \ket{\Psi_2^{\rm B}}
 =&X_{\rm B}^{\dag} \ket{0}_{\rm B},
\end{align}
where $X_{\rm A}^{\dag}$ and $X_{\rm B}^{\dag}$ denote normalized
operators that create the common parts of A and B regions,
respectively. Then we get
\begin{equation}
 \Lambda=\frac{1}{\sqrt{2}}
  \left[
   \begin{array}{ccc}
    0 & 1 \\
    1 & 0
   \end{array}
  \right]
,\quad
  \Lambda \Lambda^T=
  \frac{1}{2}
  \left[
   \begin{array}{cc}
    1 & 0\\
    0 & 1
   \end{array}
  \right].
\end{equation}
The EE is easily obtained by using the eigenvalues $\lambda_i$ of the
matrix $\Lambda\Lambda^T$ as
\begin{equation}
 S^{\rm A}=-\sum_{i}\lambda_i\log\lambda_i=\log 2.
\end{equation}
This result is for an open boundariy system where the two regions are
cut at one bond.  Therefore, the EE for the periodic boundariy system is
$S^{\rm A}=\log 4$.  These results can also be obtained by using the
matrix product representation of the wave
function~\cite{Itoh-N-M,Nakamura-O-I}.

For the PN state in the Kagom\'e lattice with $1/3$-filling, we consider
a case that two regions A and B are connected via a single triangle, for
simplicity, as shown in Fig.~\ref{fig:dividing}(b).  Then
$\ket{\Psi_n^{\rm A}}$ and $\ket{\Psi_n^{\rm B}}$ are given as
\begin{align}
 \ket{\Psi_1^{\rm A}}
 =&X_{\rm A}^{\dag} c_{i\sigma}^\dag \ket{0}_{\rm A},\\
 \ket{\Psi_2^{\rm A}}
 =&X_{\rm A}^{\dag}\ket{0}_{\rm A},\\
 \ket{\Psi_1^{\rm B}}
 =&c_{j\sigma}^\dag X_{\rm B}^{\dag} \ket{0}_{\rm B},\\
 \ket{\Psi_2^{\rm B}}
 =&c_{k\sigma}^\dag X_{\rm B}^{\dag} \ket{0}_{\rm B},\\
 \ket{\Psi_3^{\rm B}}
 =&X_{\rm B}^{\dag} \ket{0}_{\rm B}.
\end{align}
In this case, we get the following matrix elements
\begin{equation}
 \Lambda=\frac{1}{\sqrt{3}}
  \left[
   \begin{array}{ccc}
    1 & 1& 0 \\
    0 & 0 & 1
   \end{array}
  \right]
,\quad
  \Lambda \Lambda^T=
  \frac{1}{3}
  \left[
   \begin{array}{cc}
    1 & 0\\
    0 & 2
   \end{array}
  \right].
\end{equation}
If we cut the triangle in the opposite way, we should consider the
situation A$\leftrightarrow$B. In this case the matrix in
Eq.~(\ref{RDM}) becomes
\begin{equation}
  \Lambda^T \Lambda=
  \frac{1}{3}
  \left[
   \begin{array}{ccc}
    1 & 1& 0 \\
    1 & 1& 0 \\
    0 & 0 & 1
   \end{array}
  \right].
\end{equation}
The eigenvalues of the matrix $\Lambda^T \Lambda$ are
\begin{equation}
 \lambda_i=\left\{\frac{1}{3},\frac{2}{3},0\right\}.
\end{equation}
Thus the value of the EE does not depend on the ways to cut the
triangle, so that we get the EE in general cases as
\begin{equation}
 S^{\rm A}
  =N_{\bigtriangleup}\underbrace{[\log3-(2/3)\log2]}_{s_0},
\end{equation}
where $s_0=0.636514168\cdots$ and $N_{\bigtriangleup}$ means the number
of triangles along the cutting lines. This means that the EE obeys the
area law. The EE for the PN state at $2/3$-filling is obtained as the
same value as that of $1/3$-filling via the particle-hole
transformation. For the FM state at $1/2$ filling, the EE becomes zero.

\begin{figure}
\begin{center}
\includegraphics[width=7cm]{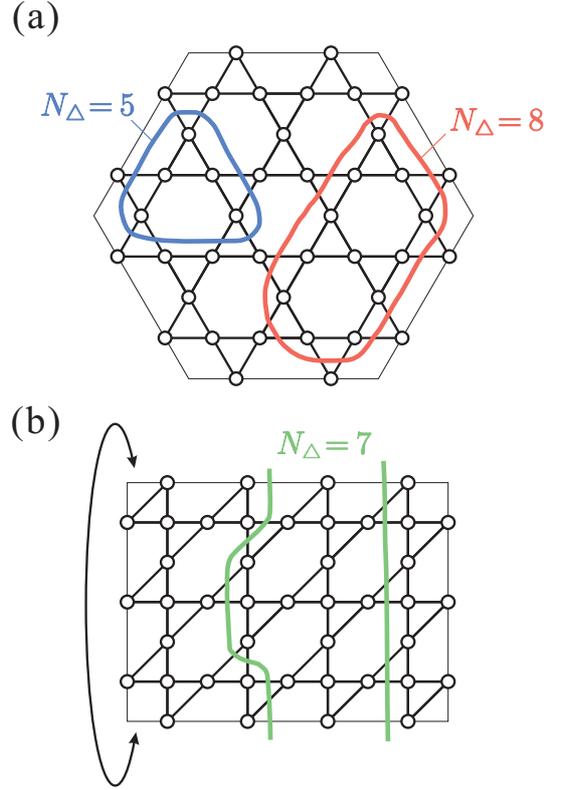}
\caption{(a) Isotropic periodic and (b) torus clusters of the Kagom\'e lattice.
The bold (cutting) lines are examples of the system division.
}
\label{fig:DMRG_EE}
\end{center}
\end{figure}

The value of EE can be easily verified numerically by using the DMRG
method.  For the BN state in 1D, the EE is $S^{\rm
A}=1.38629437\approx\log 4$ which does not depend on length of regions A
and B in a periodic chain. For the PN state in 1/3-filled Kagom\'e
lattice, some examples of the cutting lines are shown in
Fig.~\ref{fig:DMRG_EE}. We obtain $S^{\rm A}=3.182570841\approx5s_0$ and
$S^{\rm A}=5.092113346\approx8s_0$ for the periodic cluster in
Fig.~\ref{fig:DMRG_EE}(a); $S^{\rm A}=4.455599178\approx7s_0$ for the
torus cluster in Fig.~\ref{fig:DMRG_EE}(b).  Thus, we have confirmed
that the EE is proportional to the number of triangles on the cutting
lines, i.e., $S^{\rm A}=N_{\bigtriangleup}s_0$.

\section{Summary and discussion} \label{sec:summary}

In summary, we have discussed exact ground states of the generalized
Hubbard model based on the projection operator method in multicomponent
systems.  The Hamiltonian with the exact ground state can be obtained
when the lattices have bipartite structure in terms of corner sharing
unit plaquettes. We have applied this method to the 1D chain and the
Kagom\'e lattice, and obtained parameter regions of the exact ground
states for several fillings. We have also calculated the entanglement
entropy (EE). In addition, we have performed numerical calculations
based on exact diagonalization and density-matrix renormalization group,
and confirmed the results.

In the 1D chain, the exact ground state is the bond N\'eel (BN) state
where the system has a N\'eel ordered state on the
bonds~\cite{Itoh-N-M,Nakamura-O-I}. This corresponds to the staggered
dimer states in the spin-$1/2$ two-leg ladder model with four spin
exchanges.\cite{Kolezhuk-M} We have numerically confirmed the existence
of the exact BN ground state.  The BN phase may be expanded to the
outside of the analytical argument. The ferromagnetic (FM) and BN phase
boundary agrees perfectly between the analytical and numerical results.

In the Kagom\'e lattice, we have discussed the exact plaquette N\'eel
(PN) state at 1/3-filling~\cite{Nakamura-I_2004}, and also the PN state
at 2/3-filling as well as the FM state at half-filling. According to the
numerical calculations, each the exact state seems to be stabilized in a
wider region than those suggested by the analytical result.  However,
further calculations are required to corroborate it. For the EE, we have
confirmed perfect agreement between the analytical and the numerical
calculations.

In addition to the PN state, we may introduce other exact plaquette
ground states. For example, the following state
\begin{equation}
 \ket{\Psi_{\sigma}}\equiv
 \prod_{\braket{ijk}\in\bigtriangleup} B_{ijk\sigma}^{\dag}
  \prod_{\braket{i'j'k'}\in\bigtriangledown}
  C_{i'j'k'\bar{\sigma}}^{\dag}\ket{0},
  \label{Kagome_topo_state}
\end{equation}
seems like a ``topological state'', since local spin current state with
time-reversal symmetry \cite{Kane-M}. In order to stabilize this state,
we have to extend our model Hamiltonian to include the current terms
$J_{ijk\sigma}$.

\section{Acknowledgment}

M. N. acknowledges the Visiting Researcher's Program of the Institute
for Solid State Physics, the University of Tokyo, and the Max Planck
Institute f\"ur Physik komplexer Systeme, Dresden where this work was
initiated.  M.~N. is supported by JSPS KAKENHI Grant Number 17K05580.
S.~N. acknowledges support from the SFB 1143 of the Deutsche
Forschungsgemeinschaft. S.~N. would like to thank U. Nitzsche for
technical assistance.

%
%

\pagebreak

\appendix

\section{Details of calculation}\label{details}

For the 1D case, we have used the relation,
\begin{align}
&T_{ij\uparrow}T_{ij\downarrow}
  =\underbrace{\frac{1}{2}\sum_{ij\sigma\sigma'}
  T_{ij\sigma}T_{ij\sigma'}}_{h_W}
-\frac{1}{2}\sum_{\sigma}(n_{i\sigma}+n_{j\sigma})
 +\underbrace{\sum_{\sigma}n_{i\sigma}n_{j\sigma}}_{h_{V_{\parallel}}}.
\end{align}
For the Kagom\'e lattice, products of the operators generate the
following terms,
\begin{align}
&N_{ijk\uparrow}N_{ijk\downarrow}
 =\underbrace{n_{i\uparrow}n_{i\downarrow}+n_{j\uparrow}n_{j\downarrow}
 +n_{k\uparrow}n_{k\downarrow}}_{2 h_U}\\
 &+\underbrace{n_{i\uparrow}n_{j\downarrow}+n_{j\uparrow}n_{k\downarrow}
  +n_{k\uparrow}n_{i\downarrow}
  +n_{j\uparrow}n_{i\downarrow}+n_{k\uparrow}n_{j\downarrow}
  +n_{i\uparrow}n_{k\downarrow}}_{h_{V_{\perp}}},\nonumber\\
&T_{ijk\uparrow}T_{ijk\downarrow}
  =\underbrace{\frac{1}{2}\sum_{\mu\nu\sigma\sigma'}
  T_{\mu\nu\sigma}T_{\mu\nu\sigma'}}_{h_W}
 -\sum_{\sigma}N_{ijk\sigma}\\
&\qquad\qquad\qquad+\underbrace{\sum_{\sigma}(n_{i\sigma}n_{j\sigma}
 +n_{j\sigma}n_{k\sigma}+n_{k\sigma}n_{i\sigma})}_{h_{V_{\parallel}}}
 \nonumber\\
 &\qquad\qquad\qquad+
 \underbrace{
 \sum_{\sigma}
  \left(T_{ij\sigma}T_{jk\bar{\sigma}}
 +T_{jk\sigma}T_{ki\bar{\sigma}}+T_{ki\sigma}T_{ij\bar{\sigma}}
  \right)}_{h_{W'}},\nonumber\\
&N_{ijk\uparrow}T_{ijk\downarrow}+T_{ijk\uparrow}N_{ijk\downarrow}
 =\underbrace{\sum_{\mu\nu\sigma}T_{\mu\nu\sigma}
  (n_{\mu\bar{\sigma}}+n_{\nu\bar{\sigma}})}_{h_{X}}\\
&\qquad\qquad\qquad
  +\underbrace{\sum_{\sigma}\left(
  n_{i\sigma}T_{jk\bar{\sigma}}
  +n_{j\sigma}T_{ki\bar{\sigma}}
  +n_{k\sigma}T_{ij\bar{\sigma}}\right)}_{h_{X'}},\nonumber
\end{align}
where $\sum_{\mu\nu}$ is taken for $\mu\nu\in \{ij,jk,ki\}$ in one
triangle. The weight of $h_U$ term is doubled, because the on-site
interaction is shared with the neighboring triangle.

\section{Periodic Kagom\'e clusters used in DMRG calculations}

\begin{figure}
\begin{center}
\includegraphics[width=6cm]{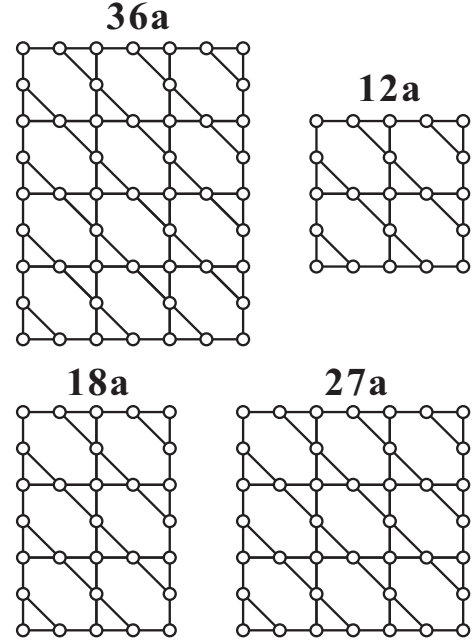}
\end{center}
\caption{Anisotropic Kagom\'e clusters.}
\label{anisocluster}
\end{figure}

\begin{figure}
\begin{center}
\includegraphics[width=7cm]{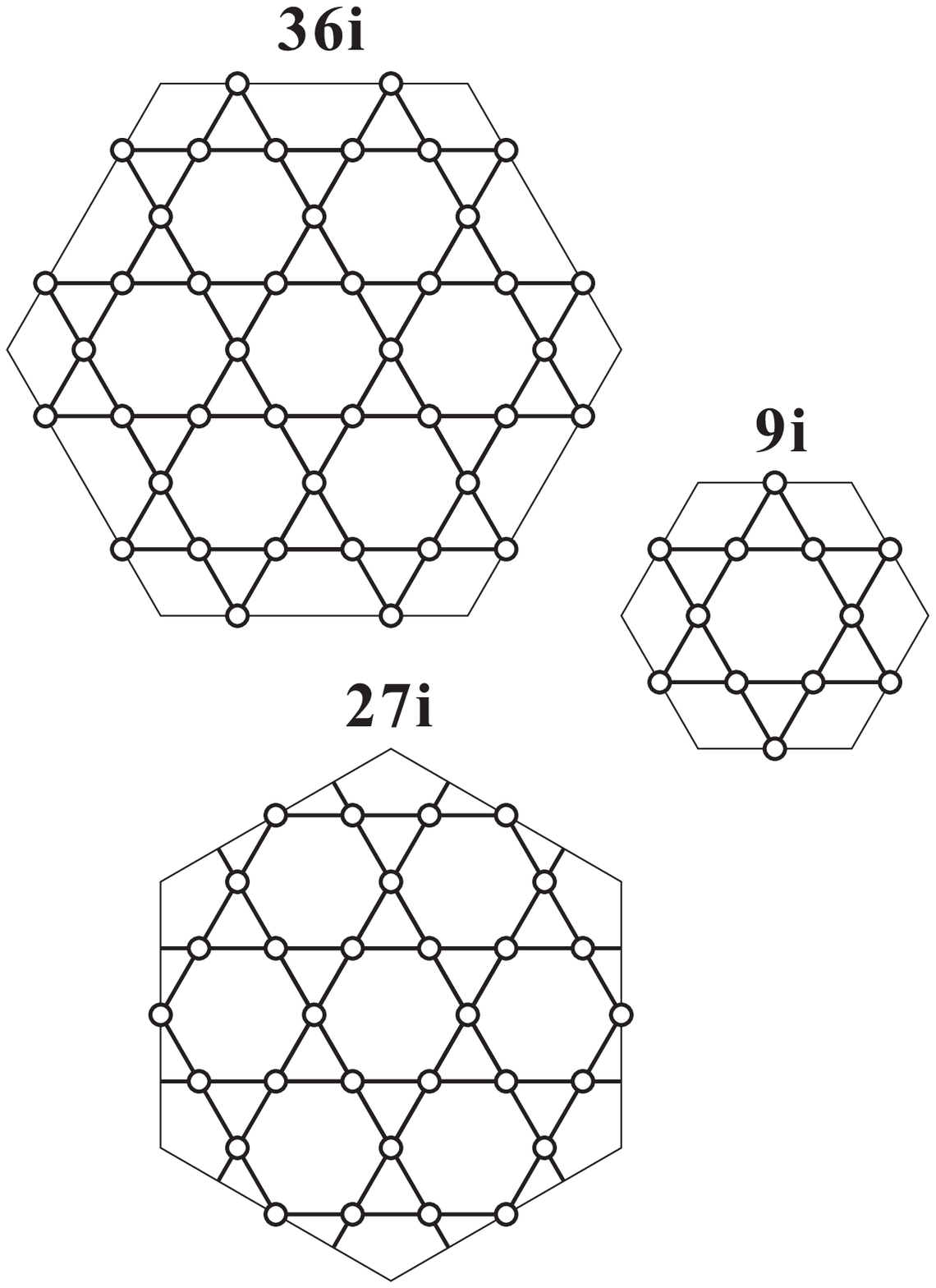}
\end{center}
\caption{Isotropic Kagom\'e clusters.}
\label{isocluster}
\end{figure}

In the exact diagonalization and density-matrix renormalization group
calculations for the Kagom\'e lattice, we used periodic clusters.  The
clusters shown in Figs.~\ref{anisocluster} and \ref{isocluster} are
spatially anisotropic and isotropic, respectively.  The periodicity of
the plaquette N\'eel state is compatible with all the clusters.

\end{document}

%% file: fig1.tex
 \begin{center}
     \psset{unit=8mm}
    \begin{pspicture}(0,-1.0)(8,5.5)
\rput(3,5){
   \multirput(0,0)(2,0){2}{
   \rput(0,0.3){
   \psline[linecolor=blue,linewidth=1.2pt](0,0)(1,0)
   \psline[linecolor=red,linewidth=1.2pt](1,0)(2,0)
   }}
   \multirput(0,0)(1,0){5}{
   \rput(0,0.3){
   \psdots[dotstyle=o](0,0)
   }}
   \rput[c](2,0){$i$}
   \rput[c](3,0){$j$}
   \rput[c](0.5,0.6){$\uparrow$}
   \rput[c](1.5,0.6){$\downarrow$}
   \rput[c](2.5,0.6){$\uparrow$}
   \rput[c](3.5,0.6){$\downarrow$}
   \rput[l](-3,0.3){(a)1D chain}
}
     \rput(3,0){
     \multirput(0,0)(2,0){2}{
     \rput(0,0){
     \pspolygon[linecolor=blue,linewidth=1.2pt]
     (0,0)(1,0)(0.5,0.8660254)
     \psline[linewidth=0.3pt]
     (0.5,0.28867513)(0,0)
     (0.5,0.28867513)(0.5,0.8660254)
     (0.5,0.28867513)(1,0)
     }
     \rput(0,1.7320508){
     \pspolygon[linecolor=red,linewidth=1.2pt]
     (0,0)(1,0)(0.5,-0.8660254)
     \psline[linewidth=0.3pt]
     (0.5,-0.28867513)(0,0)
     (0.5,-0.28867513)(0.5,-0.8660254)
     (0.5,-0.28867513)(1,0)
     }
     \rput(1,1.7320508){
     \pspolygon[linecolor=blue,linewidth=1.2pt]
     (0,0)(1,0)(0.5,0.8660254)
     \psline[linewidth=0.3pt]
     (0.5,0.28867513)(0,0)
     (0.5,0.28867513)(0.5,0.8660254)
     (0.5,0.28867513)(1,0)
     }
     \rput(1,3.4641016){
     \pspolygon[linecolor=red,linewidth=1.2pt]
     (0,0)(1,0)(0.5,-0.8660254)
     \psline[linewidth=0.3pt]
     (0.5,-0.28867513)(0,0)
     (0.5,-0.28867513)(0.5,-0.8660254)
     (0.5,-0.28867513)(1,0)
     }
     \rput(0,3.4641016){
     \pspolygon[linecolor=blue,linewidth=1.2pt]
     (0,0)(1,0)(0.5,0.8660254)
     \psline[linewidth=0.3pt]
     (0.5,0.28867513)(0,0)
     (0.5,0.28867513)(0.5,0.8660254)
     (0.5,0.28867513)(1,0)
     }
     \rput(1,0){
     \pspolygon[linecolor=red,linewidth=1.2pt]
     (0,0)(1,0)(0.5,-0.8660254)
     \psline[linewidth=0.3pt]
     (0.5,-0.28867513)(0,0)
     (0.5,-0.28867513)(0.5,-0.8660254)
     (0.5,-0.28867513)(1,0)
     }
     }
     \multirput(0,0)(2,0){2}{
     \rput(0,0){
     \psdots[dotstyle=o](0,0)(1,0)(0.5,0.8660254)
     }
     \rput(0,1.7320508){
     \psdots[dotstyle=o](0,0)(1,0)(0.5,-0.8660254)
     }
     \rput(1,1.7320508){
     \psdots[dotstyle=o](0,0)(1,0)(0.5,0.8660254)
     }
     \rput(1,3.4641016){
     \psdots[dotstyle=o](0,0)(1,0)(0.5,-0.8660254)
     }
     \rput(0,3.4641016){
     \psdots[dotstyle=o](0,0)(1,0)(0.5,0.8660254)
     }
     \rput(1,0){
     \psdots[dotstyle=o](0,0)(1,0)(0.5,-0.8660254)
     }
     }
     \rput[r](2,1.5){$j$}
     \rput[l](1,1.5){$i$}
     \rput[r](1.3,2.5){$k$}
     \rput[l](-3,1.7){(b)Kagom\'e}
     }
    \end{pspicture}
 \end{center}

%% file: fig2.tex
\begin{center}
 \psset{unit=8mm} 
 \begin{pspicture}(-5,-1.5)(5,4)
  \pspolygon*[linecolor=lightgray](4,4)(-2,1)(1,0.5)(5,2.5)(5,4)
  \pspolygon*[linecolor=lightgray](5,2.5)(1,0.5)(5,0.5)
  \pspolygon*[linecolor=lightgray](-5,2.5)(-1,0.5)(-5,0.5)
  \psgrid[subgriddiv=1,griddots=5,gridlabels=5pt](-5,-1)(5,4)
  \psaxes[labels=none,ticks=none]{->}(0,0)(-5,-1)(5,4)
  \rput[r]{0}(-0.2,3.8){$W/t$}
  \rput[r]{0}(4.9,-0.3){$U/t$}
  \psline[linewidth=0.5pt,linecolor=red]{-}(-2,1)(1,0.5)
  \psline[linewidth=0.5pt,linecolor=red]{-}(-2,1)(4,4)
  \psline[linewidth=1.0pt,linecolor=red]{-}(1,0.5)(5,2.5)
  \psline[linewidth=0.5pt,linecolor=red]{-}(1,0.5)(5,0.5)
  \psline[linewidth=0.5pt,linecolor=red]{-}(-5,2.5)(-1,0.5)(-5,0.5)
  \psline[linewidth=1.0pt,linestyle=dashed,linecolor=red]{-}%
  (1,0.5)(-0.5,-0.25)
  \psline[linewidth=0.5pt,linestyle=dashed,linecolor=red]{-}%
  (-5,1)(5,1)
  \rput[c]{26}(2.5,3.6){$\lambda_{\bar{A}\bar{A}}=0$}
  \rput[l]{26}(1.4,1){$\lambda_{\bar{A}B}=0$ (spin SU(2))}
  \rput[l]{0}(-2,-0.5){$\lambda_{BB}=0$}
  \psline[linewidth=0.5pt,linestyle=solid]{->}(-1.3,-0.3)(-0.2,0.7)
  \rput[c]{0}(2,2){\Large BN}
  \rput[l]{0}(3.8,1.4){\Large FM}
  \rput[r]{0}(-3.8,1.4){\Large PS}
  \rput[l]{0}(0.3,-0.7){$\lambda_{\bar{A}\bar{A}}=\lambda_{BB}$ (p-h symmetry)}
  \psline[linewidth=0.5pt,linestyle=solid]{->}(2.5,-0.5)(2.5,1.0)
  \rput[l]{0}(-3.5,2.2){charge SU(2)}
  \psdot[linewidth=0.5pt](-2,1.0)
  \psline[linewidth=0.5pt,linestyle=solid]{->}(-2.5,2)(-2,1.0)
 \end{pspicture}

%% file: fig4.tex
\begin{center}
  \psset{unit=12mm}
 \begin{pspicture}(2,0)(7,2)
  \multirput(0,0.8660254)(2,0){3}{
  \pspolygon[linecolor=blue,linewidth=1.2pt]
  (2,0)(3,0)(2.5,0.8660254)
  \psline[linewidth=0.1pt]
  (2.5,0.28867513)(2,0)
  (2.5,0.28867513)(2.5,0.8660254)
  (2.5,0.28867513)(3,0)
  \rput[r](1.9,0){$i$}
  \rput[l](3.1,0){$j$}
  \rput[c](2.5,1.1){$k$}
  }
  \rput(2,0.8660254){
  \psline[linecolor=blue,linewidth=1.2pt]{<-}%
  (2.6,0)(2.4,0)
  \psline[linecolor=blue,linewidth=1.2pt]{<-}%
  (2.72,0.48497423)(2.80,0.34641016)
  \psline[linecolor=blue,linewidth=1.2pt]{<-}%
  (2.20,0.34641016)(2.29,0.50229473)
  }
  \rput(4,0.8660254){
  \psline[linecolor=blue,linewidth=1.2pt]{->}%
  (2.6,0)(2.4,0)
  \psline[linecolor=blue,linewidth=1.2pt]{->}%
  (2.72,0.48497423)(2.80,0.34641016)
  \psline[linecolor=blue,linewidth=1.2pt]{->}%
  (2.21,0.36373067)(2.29,0.50229473)
  }
  \rput[c](2.5,0.2){$\ket{A_{ijk\sigma}}$}
  \rput[c](4.5,0.2){$\ket{B_{ijk\sigma}}$}
  \rput[c](6.5,0.2){$\ket{C_{ijk\sigma}}$}
 \end{pspicture}
\end{center}

%% file: fig5.tex
\begin{center} 
 \psset{unit=8mm}
 \begin{pspicture}(-3,-2.0)(5.5,4)
  \pspolygon*[linecolor=lightgray]
  (-2.6666667,0.6666666)(0.66666667,0.3333333)(5.0,0.875)(5,1.625)
  \psgrid[subgriddiv=1,griddots=5,gridlabels=7pt](-3,-1)(5,4)
  \psaxes[labels=none,ticks=none]{->}(0,0)(-3,-1)(5,4)
  \rput[l]{0}(0.1,3.8){$W/t$}
  \rput[r]{0}(5,-0.3){{$U/t$}}
  \psline[linecolor=red]{-}(-3,0.625)(5,1.625)
  \psline[linecolor=red]{-}(-3,-0.125)(5.0,0.875)
  \psline[linecolor=red]{-}(-3,0.7)(5,-0.1)
  \rput[c]{0}(2.0,0.85){\Large PN}
  \rput[c]{7}(2,1.5){$\lambda_{\bar{A}\bar{A}}=0$}
  \rput[c]{7}(3.5,0.4){$\lambda_{\bar{A}B}=0$}
  \rput[c]{-6}(-1.9,0.4){$\lambda_{BB}=0$}
  \rput[l]{0}(-3,3.7){\large (a) $t>0$}
 \end{pspicture}
 \begin{pspicture}(-3,-1.5)(5.5,4)
  \pspolygon*[linecolor=lightgray]
  (-1.3333333,0.3333333)(1.3333333,0.16666667)(5,2)(5,3.5)
  \psgrid[subgriddiv=1,griddots=5,gridlabels=7pt](-3,-1)(5,4)
  \psaxes[labels=none,ticks=none]{->}(0,0)(-3,-1)(5,4)
  \rput[l]{0}(0.1,3.8){$W/|t|$}
  \rput[r]{0}(5,-0.3){{$U/|t|$}}
  \psplot[plotpoints=50,plotstyle=curve,linecolor=red]%
  {-3}{5}{x -16 div 0.25 add}
  \psline[linecolor=red]{-}(-3,-0.5)(5,3.5)
  \psline[linecolor=red]{-}(-1,-1)(5,2)
  \rput[c]{0}(2.0,1.35){\Large PN}
  \rput[c]{26}(2,2.3){$\lambda_{\bar{B}\bar{B}}=0$}
  \rput[c]{26}(3.5,1.0){$\lambda_{A\bar{B}}=0$}
  \rput[c]{-5}(-2.1,0.6){$\lambda_{AA}=0$}
  \rput[l]{0}(-3,3.7){\large (b) $t<0$}
 \end{pspicture}
\end{center}

%% file: fig8.tex
\begin{center} 
 \psset{unit=8mm}
 \begin{pspicture}(-1.5,-2.5)(7.5,2)
  \pspolygon*[linecolor=lightgray]
  (7,1.91666667)(0.66666667,0.33333333)
  (2.6666667,-0.66666667)(7,-1.20833333)
  \psgrid[subgriddiv=1,griddots=5,gridlabels=7pt](-1,-2)(7,2)
  \psaxes[labels=none,ticks=none]{->}(0,0)(-1,-2)(7,2)
  \rput[r]{0}(-0.1,1.8){$W/t$}  
  \rput[r]{0}(7,-0.3){{$U/t$}}
  \psplot[plotpoints=50,plotstyle=curve,linecolor=red]%
  {-1}{7}{x 0.25 mul 0.1666667 add}
  \psplot[plotpoints=50,plotstyle=curve,linecolor=red]%
  {-1}{7}{x -0.125 mul 0.333333 sub}
  \psplot[plotpoints=50,plotstyle=curve,linecolor=red]%
  {-1}{4}{x -0.5 mul 0.6666667 add}
  \rput[c]{0}(4.5,0.35){\Large FM}
  \rput[c]{15}(4,1.4){$\lambda_{\bar{A}\bar{B}}=0$}
  \rput[c]{-7}(5,-1.2){$\lambda_{\bar{A}\bar{A}}=0$}
  \rput[c]{-25}(-0.4,0.6){$\lambda_{\bar{B}\bar{B}}=0$}
  \rput[l]{0}(0.2,-1.7){\large (a) $t>0$}
 \end{pspicture}
 \begin{pspicture}(-1.5,-3.0)(7.5,2)
  \pspolygon*[linecolor=lightgray]
  (7,1.5833333)(7,-0.541666667)(1.3333333,0.16666667)
  \psgrid[subgriddiv=1,griddots=5,gridlabels=7pt](-1,-2)(7,2)
  \psaxes[labels=none,ticks=none]{->}(0,0)(-1,-2)(7,2)
  \rput[r]{0}(-0.1,1.8){$W/|t|$}  
  \rput[r]{0}(7,-0.3){{$U/|t|$}}
  \psplot[plotpoints=50,plotstyle=curve,linecolor=red]%
  {-1}{7}{x 0.25 mul 0.1666667 sub}
  \psplot[plotpoints=50,plotstyle=curve,linecolor=red]%
  {-1}{7}{x -0.125 mul 0.333333 add}
  \rput[c]{0}(4.5,0.35){\Large FM}
  \rput[c]{15}(4,1.07){$\lambda_{\bar{A}\bar{B}}=0$}
  \rput[c]{-7}(5,-0.55){$\lambda_{\bar{A}\bar{A}}=0$}
  \rput[l]{0}(0.2,-1.7){\large (b) $t<0$}
 \end{pspicture}
\end{center}

%% file: fig10.tex
 \begin{center}
  \psset{unit=8mm}
  \begin{pspicture}(0,1.0)(8,4.5)
\rput(3,4){
   \multirput(0,0)(2,0){2}{
   \rput(0,0.3){
   \psline[linecolor=blue,linewidth=1.2pt](0,0)(1,0)
   \psline[linecolor=red,linewidth=1.2pt](1,0)(2,0)
   }}
   \psline[linecolor=blue,linewidth=1.2pt](4,0.3)(5,0.3)
   \multirput(0,0)(1,0){6}{
   \rput(0,0.3){
   \psdots[dotstyle=o](0,0)
   }}
   \rput[c](2,0){$i$}
   \rput[c](3,0){$j$}
   \rput[c](0.4,0.6){$\uparrow$}
   \rput[c](1.4,0.6){$\downarrow$}
   \rput[c](2.4,0.6){$\uparrow$}
   \rput[c](3.4,0.6){$\downarrow$}
   \rput[c](4.4,0.6){$\uparrow$}
   \rput[c](1,-0.1){A}
   \rput[c](4,-0.1){B}
   \rput[l](-3,0.3){(a)1D chain}
   \psline[linecolor=PineGreen,linestyle=dashed,linewidth=1.2pt]%
   (2.5,0)(2.5,0.6)
}
\rput(3.5,2){
   \rput[l](-3.5,0){(b)Kagom\'e}
   \rput{-30}(0,0){
   \pspolygon[linecolor=blue,linewidth=1.2pt]
   (0,0)(1,0)(0.5,0.8660254)
   \psline[linewidth=0.3pt]
   (0.5,0.28867513)(0,0)
   (0.5,0.28867513)(0.5,0.8660254)
   (0.5,0.28867513)(1,0)
   \psdots[dotstyle=o](0,0)(1,0)(0.5,0.8660254)
   }
   \rput[c](-0.2,0){$i$}
   \rput[l](1.0,0.5){$j$}
   \rput[l](1.0,-0.5){$k$}
   \psline[linecolor=PineGreen,linestyle=dashed,linewidth=1.2pt]%
   (0.5,-1.0)(0.5,0.6)
   \rput[c](0.2,-0.9){A}
   \rput[c](0.8,-0.9){B}
   \rput{30}(3,-0.5){
   \pspolygon[linecolor=red,linewidth=1.2pt]
   (0,0)(1,0)(0.5,0.8660254)
   \psline[linewidth=0.3pt]
   (0.5,0.28867513)(0,0)
   (0.5,0.28867513)(0.5,0.8660254)
   (0.5,0.28867513)(1,0)
   \psdots[dotstyle=o](0,0)(1,0)(0.5,0.8660254)
   }
   \rput[c](4.1,0){$i$}
   \rput[l](2.65,0.5){$j$}
   \rput[l](2.65,-0.5){$k$}
   \psline[linecolor=PineGreen,linestyle=dashed,linewidth=1.2pt]%
   (3.35,-1.0)(3.35,0.6)
   \rput[c](3.05,-0.9){A}
   \rput[c](3.65,-0.9){B}
}
  \end{pspicture}
 \end{center}